\documentclass[runningheads]{llncs}
\usepackage{graphicx}
\usepackage{geometry}
\usepackage{rotating}
\usepackage{tabularx}
\usepackage{makecell}
\usepackage{multirow}
\usepackage{siunitx}
\usepackage{hyperref}
\usepackage{graphicx}
\usepackage{subcaption}
\usepackage{float}
\usepackage{wrapfig}
\usepackage{realboxes}
\usepackage{setspace}
\usepackage{hhline}
\usepackage{algorithm}
\usepackage{algpseudocode}
\usepackage{booktabs}
\usepackage{amsmath}
\usepackage{scrlayer-scrpage}
\usepackage{caption}
\usepackage{multirow}
\usepackage{graphicx}
\usepackage[table]{xcolor}
\usepackage{inputenc} % allow utf-8 input
\usepackage{fontenc}    % use 8-bit T1 fonts
\usepackage{hyperref}       % hyperlinks
\usepackage{url}            % simple URL typesetting
\usepackage{booktabs}       % professional-quality tables
\usepackage{amsfonts}       % blackboard math symbols
\usepackage{nicefrac}       % compact symbols for 1/2, etc.
\usepackage{microtype}      % microtypography
\usepackage[backend=biber,style=numeric, sorting=nyt]{biblatex} % Use the bibtex backend with the authoryear citation style (which resembles APA)

\usepackage[normalem]{ulem}
\usepackage{color,soul}

\begin{document}
\title{On the Power of Graph Neural Networks and Feature Augmentation Strategies to Classify Social Networks}
%\thanks{Supported by X.}
%
\titlerunning{On the Power of GNNs}
\author{Walid Guettala%\inst{1}
\orcidID{0000-0001-5941-6818} \and
László Gulyás%\inst{1}
\orcidID{0000-0002-6367-6695}}
\authorrunning{Guettala and Gulyás}
\institute{ELTE Eötvös Loránd University, Budapest, Hungary;  
Department of Artificial Intelligence\\
\email{\{guettalawalid,lgulyas\}@inf.elte.hu}}
\maketitle              % typeset the header of the contribution
\begin{abstract}
This paper studies four Graph Neural Network architectures (GNNs) for a graph classification task on a  synthetic dataset created using classic generative models of Network Science. Since the  synthetic networks do not contain (node or edge) features, five different augmentation strategies (artificial feature types) are applied to nodes. All combinations of the 4 GNNs (GCN with Hierarchical and Global aggregation, GIN and GATv2) and the 5 feature types (constant 1, noise, degree, normalized degree and ID -- a vector of the number of cycles of various lengths) are studied and their performances compared as a function of the hidden dimension of artificial neural networks used in the GNNs. The generalisation ability of these models is also analysed using a second synthetic network dataset (containing networks of different sizes).  
%The paper delves into the realm of Graph Neural Networks (GNNs), analyzing their efficacy and adaptability through a detailed study of four distinct architectures—GIN, Hierarchical, Global, and GATv2—across varied node feature types and hidden dimension sizes. This investigation aims at comprehending model accuracy and generalization capabilities. 
%The dataset, meticulously crafted to mirror real-world networks, encapsulates fundamental graph properties and well-known model types, aiming to cover all possible configurations of four key properties, fostering a total of 8 configurations. 
%Notably, the dataset's design specifically explores scenarios with contrasting transitivity and average shortest paths, incorporating scale-free graph models for degree distribution. Segmented into small and medium datasets, this resource offers 2000 samples, enabling comprehensive training, validation, and testing, with a particular focus on medium network sizes to assess model generalization. 
%Further enriching the dataset understanding, four feature types—Ones, Noise, Degree, and Identity Features—were studied, shedding light on their varied impacts on performance. The findings underscore the significance of adept architecture selection and feature integration in enhancing model discriminative capabilities within graph-based data representations. 
Our results point towards the balanced importance of the computational power of the GNN architecture and the the information level provided by the artificial features. GNN architectures with higher computational power, like GIN and GATv2, perform well for most augmentation strategies. On the other hand, artificial features with higher information content, like ID or degree, not only consistently outperform other augmentation strategies, but can also help GNN architectures with lower computational power to achieve good performance. 
%The investigation reveals the consistent superiority of the identity feature in achieving higher accuracy and robust generalization across architectures, emphasizing the pivotal role of informed feature selection in optimizing GNNs.

\end{abstract}

%-----------------------------------------------------------------------------------
\section{Introduction}

The usage of Graph Neural Networks (GNNs), i.e., deep learning approaches to the embedding of graphs, is gaining momentum in many fields of research and applications. These include the field of complex, or social network analysis, the domain of studying the structure of  real-world networks with the goal of gaining insights about the behavior of the system that they represent. This is done by the calculation of various measures (network statistics) at the level of nodes, groups of nodes or the entire network. With the plethora of different possible statistics to calculate, an alternative approach is to compare the network of study to abstract families of networks from Network Science with well known structural properties. These classes of networks are the products of stochastic generative models that were designed to replicate some of of the measurable characteristics that are common in real-world networks. For example, Barabási-Albert networks have scale-free degree distributions \cite{BA}, while Watts-Strogatz networks have a combination of high transitivity and short average path length \cite{WS}. This approach, i.e., the comparison of the network in question to some well-known network types, is, in essence, similar to the task of \textit{graph classification}. The task is to assess the similarity of the network in question to the abstract network families and determine the likelihood of it belonging to the class of that family.

In this paper we embark on a study of various deep learning graph embedding approaches applied to the task of social network analysis. To this end, we first create a dataset containing a balanced number of instances from classic generative network models with carefully chosen parameter combinations. This dataset is then used to train a select set of graph classification models, using a variety of graph neural network approaches with a wide array of hyperparameter combinations. The trained models are then tested on both unseen samples from the original dataset and on synthetic networks generated using different parameter combinations. The accuracy of the models for the two types of tests are then compared and its dependence on the various hyperparameters is analysed. 
%The application of the best models to real-world social networks is also demonstrated.

The contributions of this paper are as follows. A dataset generated from classic generative models of Network Science is introduced. This dataset is used to train graph classification models with the aim to characterise real-world networks. The behavior of a select set of graph neural network approaches (GNNs) is studied and compared on the task of graph characterisation. In addition 5 augmentation strategies are also studied to augment nodes with artificial features to provide GNNs with structural information about the networks. Our results point towards the balanced importance of the computational power of the GNN architecture and the the information level provided by the artificial features.

\section{Related works}

%\subsection{Graph Neural Networks (GNNs) Overview}
Graph Neural Networks (GNNs) are designed to create node (and thus graph) embeddings using neural network mechanisms tailored to graph structures \cite{GCN, gao2019graph}. These architectures try to encode the nodes' structural position within the network by implementing information exchange among neighboring nodes (i.e., via message passing). This approach allows node  representations to capture information from their local context. 
Various architectures like graph convolutions and attention mechanisms \cite{zhang2018image} were proposed that differ in their approaches to information exchange, offering diverse embeddings that serve multiple graph mining tasks such as node classification, link prediction, and graph classification \cite{GCN, hwang2022ahp, zhang2018end}.
%\subsection{Graph Classification}
In graph categorization and classification, the challenge lies in   merging node embeddings effectively  into a comprehensive graph representation. Approaches like global pooling (e.g., mean and maximum aggregation) and hierarchical pooling (reducing nodes into supernodes) are widely used \cite{duvenaud2015convolutional, ma2019graph}. %\subsection{Identity-Aware Graph Neural Networks}
A special approach, termed identity-aware graph neural networks (ID-GNNs) was proposed in \cite{ID-GNN}, to improve GNNs prediction accuracy for nodes, edges and various graph properties. The proposal was tested on synthetic networks, similarly to our present study, so far with limited effectiveness on real-world networks.

%showcasing two setups - ID-GNN Full and ID-GNN Fast. These networks have been tested on various datasets, highlighting improvements in predicting node, edge, and graph properties, particularly on synthetic graphs. However, their effectiveness on real-world datasets remains limited, emphasizing the importance of not fixating on specific properties during model learning.

%\subsection{Benchmarking and Model Evaluations}

Together with the plethora of proposed GNN approaches the demand for benchmark comparisons has also emerged. \cite{Benchmark_dataset} studies the link prediction task on a dataset of  of 100+ real-world graphs. In contrast, our study focues on the task of graph classification and uses synthetic datasets. 
%\hl{\textbf{WALID: What do you mean by the following sentences?:} 
%emphasizing varied structural properties such as power-law degree distribution, community structure, and dense areas with sparse connections. However, it lacks in-depth analysis of properties like degree distribution variations and transitivity levels, limiting the benchmark's scope. }
%\hlc[red]{\textbf{Prof: I mean the structural properties on which their dataset focuses, but they say it is too vague. They don't know specifically which type of graph possesses these properties and which ones do not. Consequently, they are unable to discern when the models perform well or poorly in relation to these properties.]}}
%In the realm of benchmarking for link prediction, 
%In contrast, our focus on graph classification differs from their link prediction emphasis, although both explore node embedding learning and the impact of graph sizes. 
On the other hand, \cite{errica2019fair} concentrates on experimental reproducibility and fairness of architecture training, re-evaluating popular graph neural network models across 9 datasets (4 chemical, 5 social). They emphasize the importance of consistent comparison practices and demonstrate the enhancement by adding structural features to node representations. However, their study lacks comprehensive benchmarks for model %efficacy 
assessment and  uses  real-world datasets, in contrast to our approach.
\section{Methods}

This section is structured as follows. Section \ref{sec:ExperimentSetup} introduces the setup of our experiments. This is followed by the description of the graph neural network architectures studied in Section \ref{sec:GNNsStudied}. Finally, Section \ref{sec:dataset} introduces the synthetic datasets, created using abstract generative models of Network Science. These datasets are used for training and testing purposes. The section also explains the considerations along which they were created. 

\begin{figure}[ht]
    \vspace{-1em}
    \centering
    \includegraphics[width=0.9\textwidth]{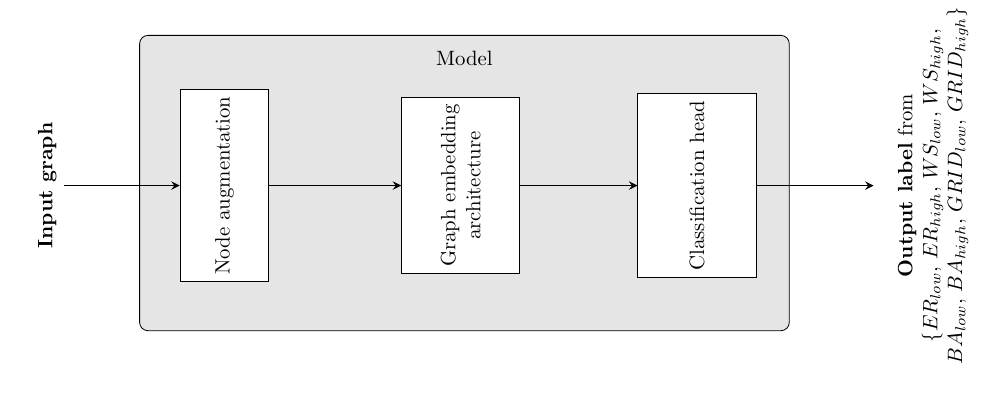}
    \caption{The generic structure of the models studied. The input graphs have no feature information, so nodes are augmented by artificial features (varying across the studies). The resulting network is passed through one of the recent graph embedding architectures (GNNs, also varying across the studies) and then classified into one of the 8 classes defined in Section \ref{sec:dataset}.}
    \label{fig:ExpSetup}
    \vspace{-2.5em}
\end{figure}

\subsection{Experiment Setup}\label{sec:ExperimentSetup}
%\vspace{-0.5em}

We studied a set of GNN models on a special graph classiciation task, using the synthetic datasets described in Section \ref{sec:dataset}. The  task took networks (graphs) without node or edge features as input and sorted them into one of the eight classes defined in Section \ref{sec:dataset}, based on the network structure alone.\footnote{The code for the experiments, together with the datasets used, is available from https://github.com/ walidgeuttala/Synthetic-Benchmark-for-Graph-Classification.}

Figure \ref{fig:ExpSetup} depicts the generic model structure. Since our abstract networks contain no feature information, the first step augments the nodes with artificial features. This is followed by the application of one of the recent graph embedding architectures. Finally, a classification (prediction) head is used to label the embedded graph by one of the eight labels. 

We experimented with 5 augmentation strategies (i.e., 5 artificial feature types) and 4 embedding architectures (GNNs). The prediction head was the same for all experiments: a fully connected artificial neural network with a 3 hidden layers and 8 output nodes, using log softmax activation. The number of nodes in the hidden layers (in all  ANNs used in the embedding architectures, as well as in the classification head), denoted by $H$, was varied in the experiments. The following values were tested: 1, 2, 3, 8, 16 and 32%, 64 and 128
.

%\hl{\textbf{WALID: Please, check the description of the prediction head and correct it if necessary. Once done, please, comment out this request.}}

For the training of each model, the loss function of negative log likelihood was used, with a minibatch size of 100 and a dropout rate of 0.5. The optimization was carried out using the ADAM  optimizer for 100 epochs. For each configuration of feature type and embedding architecture, 5 replications were carried out and the average performance was reported.

%\hl{\textbf{WALID: Please, explain the loss function you mention here. What is its official name? nll\_loss sounds like a name of a function in python, not a name of a loss function. Can you add a reference? Also, as a note to ourselves, I wonder if the batchsize of 100 is a perfect choice. Additional note: two replications is extremely low. It is almost like a single one. For the next paper, we need more data.}}

% Loss Function: nll\_loss, Number of Trials: 2, 

\subsection{Augmentation Strategies (Artificial Feature Types)}\label{sec:augmentation}

As we are interested in studying GNNs ability to learn network structure, the input graphs of our experiments contain no node or edge features. This makes their embedding difficult as nodes are indistinguishable from each other. To counter this the nodes are augmented with artificial features. 5 different augmentation strategies (i.e., feature types) are studied in our experiments (see Table \ref{tab:feature-summary}).

The simplest augmentation strategy assigns the same constant feature (the scalar of 1 for the sake of simplicity) to each node. This helps differentiating nodes with different number of neighbors and also serves as a baseline for more advanced strategies. 
A slightly more advanced strategy still assigns a scalar value to nodes, but selects the values \textit{randomly}  (from the interval of $[0, 1]$). While random values contain no structural information, they help to distinguish nodes and neighborhoods.
The third strategy introduces node dependent features by adding the degree of the node (i.e., the number of its links) as a feature. The next strategy extends on this by normalising the degree information. That is, instead of the number of neighbors, it assigns the fraction of possible links that exists. This makes the feature information independent of network size. Finally, the fifth augmentation strategy adds \textit{node identity} information as defined in \cite{ID-GNN}. This contains the number of cycles of lengths 1 to $k$ that the node is part of, whereas a cycle of length 1 is interpreted as the degree of the node. Obviously, this artificial feature is the richest in information and thus the most expensive to use. Its computational cost scales linearly with the number of nodes and edges ($\mathcal{O}(k \cdot (V + E)$, where $V$ and $E$ stand for the number of nodes and edges, respectively).

%\hl{\textbf{WALID: Technical Remark: Please, add a more readable citation label. Citing '2' is very obscure. We will never remember what '2' refers to. Add a label that either concernts the authors or the title of the paper, or something.}}

%We have enhanced the small and medium-sized graph datasets by incorporating 5 different types of features. These features were added to analyze their effect on performance and to gain insights into the limitations associated with each feature type. For more information about the specific features chosen, please refer to Table \ref{tab:feature-summary}.

\begin{small}
\begin{table}[h!]
\vspace{-1em}
\centering
\caption{Summary of Feature Types}
\label{tab:feature-summary}
\begin{tabular}{|c|p{10cm}|}
\hline
\textbf{Feature Type} & 
\multicolumn{1}{|c|}{\textbf{Description}} \\ \hline
Ones & The scalar constant of 1 is assigned to each node. \\ 
\hline
Noise & Each node is given a random number between 0 and 1, generated from a uniform distribution. 
%& \hl{\textbf{WALID: This is IMPORTANT! Are you sure that you are assigning random values from a \textit{normal} distribution?? I mean, if the distribution is normal, you can never be sure that the values are between 0 and 1. Are you sure that you are not using a \textit{uniform} distribution? There you would sample from a fixed interval with equal probability for each value.}}
\\ 
\hline
Degree & The degree of the node is added as a feature.\\ 
\hline
Norm Degree & The normalised node feature represents the ratio of the degree of each node to the maximum degree in the entire graph dataset. This is done to address the issue of varying summation pooling ranges that arise from graphs of varied sizes, which is encountered in the standard degree feature.
%& \hl{\textbf{WALID: Again, are you sure? Are you normalizing by the \textit{highest degree in the graph} as you write, or with the highest \textit{possible} degree. It is the second that you should do as that is what ensures independence across networks and network sizes. If you normalize by the actual maximum then you introduce dependence on the actual instance, which contradicts to the whole theoretical basis of normalisation.}}
\\ \hline
Identity & Adds \textit{identity information} to nodes as defined in \cite{ID-GNN}. This includes the degree, plus the number of cycles of lengths 2 to $k$ that the node is part of.\\ 
\hline
\end{tabular}
\end{table}
\end{small}

%--------------------------------------
\vspace{-2em}
\subsection{GNN Models Used}\label{sec:GNNsStudied}

\begin{figure}[ht!]
    \centering
    \includegraphics[width=0.28\textwidth]{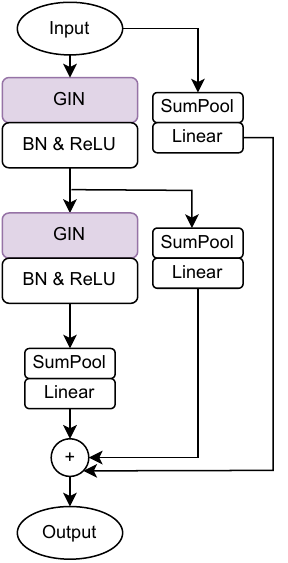}
    \includegraphics[width=0.24\textwidth]{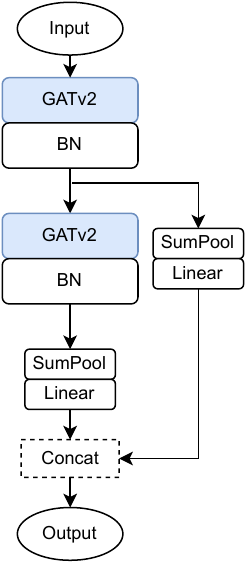}
    \includegraphics[width=0.24\textwidth]{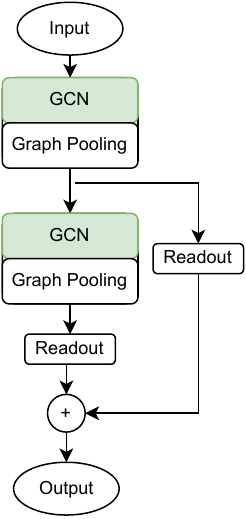}
    \includegraphics[width=0.18\textwidth]{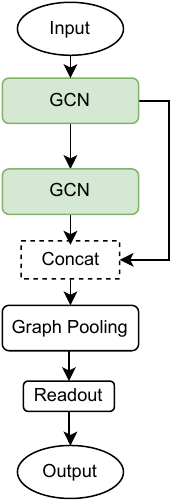}
    
        \caption{The high-level architectures of the 4 graph embedding architectures studied. GIN \cite{GIN} is on the far left, followed  by  GATv2 \cite{GATv2}. The Hierarchical and Global architectures are on the right (in this order). Each architecture is depicted with $K=2$ layers, where $K$ is a hyperparameter that is also explored in the study. The output of the embedding is to be fed to the classification head. \textit{BN} denotes Batch normalization, \textit{Readout} stands for the output of the pooling, while \textit{Linear} means a single layer of neurons). The meaning of \textit{GCN, }\textit{GIN}, \textit{GATv2} layers and \textit{Graph Pooling} is described in the text.}
    \label{fig:models}
\end{figure}

Figure \ref{fig:models} shows the four deep learning based graph embedding architectures (or GNNs) considered. These include traditional ones like GCN \cite{GCN} with Graph Pooling, but also more recent proposals like GATv2 \cite{GATv2} (GAT) and GIN \cite{GIN}. This selection, while arbitrary, encompasses  message passing methods, with various pooling techniques, as well as attention based approaches.  
%\hlc[green]{WALID: The ordering of the figures is wrong. REmember, I asked you to break up the figure so that we can move around the individual architectures? This is why... Now, how can I change the order??}

%\subsubsection{Hierarchical and Global}
%The Hierarchical and Global architectures, depicted on the right side of Figure \ref{fig:models}, both utilize 
Graph Convolutional Network (GCN) layers, as discussed in \cite{GCN} implement a message passing scheme to embed nodes based on the information collected from their local neighborhood. Their key governing parameter, $K$, controls the size of the neighborhood: the number of times nodes pass messages to their immediate neighbors (i.e., the number of 'layers' around the node from which information is collected). The messages collected from neighbors are aggregated at each node and passed on in the next message passing iteration. 
%The input comprises node representations ${ h_i \in \mathbb{R}^d \mid i \in V }$ and the set of edges $E$, resulting in a new set of node representations ${ h'_{i} \in \mathbb{R}^{d'} \mid i \in V }$ through the application of a parametric function to each node, considering its neighbors $N_i = { j \in V \mid (j, i) \in E }$.
%\[
%h'_i = f_{\theta}(h_i, \text{AGGREGATE}(\{h_j \mid j \in N_i\}))
%\]
%The distinct designs of functions $f$ and the AGGREGATE method distinguish between different types of Graph Neural Networks (GNNs). For instance, GCN utilizes an element-wise sum for AGGREGATE.
The various GCN models differ in the methods they use for aggregation and for creating messages from the aggregated information. For instance, classic GCN aggregates by the element-wise summing of messages messages received from neighbors. Figure \ref{fig:models} presents the 4 architectures studied in this paper.

The \textbf{Hierarchical} applies  Graph Pooling for aggregation, using the SAGPool method\cite{SagePool}.  SAGPool takes a graph and its node embeddings and generates a subgraph with the corresponding node embeddings. The size of the  output  is controlled by a ratio ($r \in [0,1]$) variable. The resulting readouts, i.e., the output of node pooling and summations, are subsequently passed to the next GCN layer. The iterative ($K$ times) usage of Graph Pooling gradually reduces the number of nodes. In contrast, the \textbf{Global} architecture  applies Graph Pooling only once, to the concatenated results of the $K$ consecutive GCN layers.

SAGPool \cite{SagePool} calculates attention scores for each node using a one dimensional GCN with  \textit{tanh} activation. Nodes with the highest score are selected for the output subgraph (together with their information from the application of the GCN layer prior to SAGPool). Both average and max pooling is applied to the selected nodes, yielding a combined a readout vector of twice the size of the number of nodes in the subgraph.

%\subsubsection{GIN Architecture}
The \textbf{GIN} layer was proposed in \cite{GIN} to maximize the representational power of graph embeddings, i.e., their ability to generate similar embeddings for graphs of the same structure (especially for isomorphic graphs). \cite{GIN} augments nodes with the Identity feature (see Section \ref{sec:augmentation}) and shows that a  \textit{multi-layer perceptrons (MLP)} of two layers is  required to process this information, instead of the single \textit{Linear} layer of classic GCNs.
%One way to measure a GNN's 'power' is by using the Weisfeiler-Lehman (WL) graph isomorphism test \cite{weisfeiler1968reduction}. Isomorphic graphs share the same structure: identical connections but with a permutation of nodes. GIN extends the WL test by learning to embed subtrees (node computation graph of the message passing) into a low-dimensional space. This allows GNNs not only to differentiate between different structures but also to learn how to map similar graph structures to similar embeddings. This process captures dependencies between graph structures, achieved using multi-layer perceptrons (MLPs) for each GIN layer (Basically GIN layer is a GCN layer with MLP instead of one linear layer).
As shown on Figure \ref{fig:models} (leftmost panel) the \textbf{GIN architecture} applies batch normalization (BN) and the Rectified Linear Unit (ReLU) activation function after each GIN layer. Summation Pooling, combined with a single Linear layer, is applied to the output of the GIN layers (as well as to the input) and summed element-wise to produce the final embedding.

The \textbf{GATv2 layer }enhances the graph attention mechanism introduced by Graph Attention Networks (GAT) \cite{velivckovic2017graph}. GAT itself computes a weighted average of neighboring nodes' representations, determining each node's new representation based on these weighted values. However, GAT has a limitation: it uses static attention, meaning it always gives certain nodes more weight regardless of the query 
%(i.e., node
%refers to the specific information or context a GATv2 layer uses to weigh the importance of neighboring nodes' representations
%). 
%This limitation restricts GAT's ability to adapt to different queries and contexts. 
GATv2 \cite{GATv2} resolves this by  calculating the score for each node pair differently, allowing it to achieve dynamic attention. 
%This adjustment enables GATv2 to select different nodes based on the query, enhancing its expressiveness compared to GAT. 
The \textbf{GATv2 architecture} (second left on Figure \ref{fig:models}) uses $K$ GATv2 layers, followed by batch normalisation (BN) and Summation Pooling followed by a single Linear layer, just like in the GIN architecture. However, here the results are concatenated (and not element-wise summed) to produce the embedding.

\begin{table}[h!] 
    \vspace{-1.5em}
    \centering
    \caption{The 8 network types included in the synthetic datasets and their main features. \textit{Label} identifies the generating model (ER=Erdős-Rényi, WS=Watts-Strogatz, BA=Barabási-Albert and GRID=regular lattice) with the number of links as a postfix (corresponding to the value of average degree). \textit{Degree distribution} is either scale-free (SF) or non-scale-free (NSF), while $\ell$ and $T$ stand for average path length and transitivity, respectively.}
    \label{tab:networks}
    \begin{tabular}{cccccc}
        \toprule
        \textbf{Label} & \textbf{Degree Distribution} & \hspace{1em} \textbf{$\ell$} \hspace{1em} & \textbf{$T$} & \textbf{Avg Degree} \\
        \midrule
        $ER_{low}$  & NSF & Low & Low & 4 \\
        $ER_{high}$ & NSF  & Low & Low & 8 \\
        $WS_{low}$  & NSF & Low & High & 4 \\
        $WS_{high}$ & NSF & Low & High & 8 \\
        $BA_{low}$  & SF & Low & Low & 4 \\
        $BA_{high}$ & SF & Low & Low & 8 \\
        $GRID_{low}$  & NSF & High & Low & 4 \\
        $GRID_{high}$ & NSF & High & High & 8 \\
        \bottomrule
    \end{tabular}
\end{table}

\vspace{-2em}
\subsection{Synthetic Network Datasets }\label{sec:dataset}

Two datasets were created, both containing synthetic networks produced by abstract generative models from Network Science. The first was both used to train the studied GNN models and to test their performance (on unseen samples). The second was only used for testing the generalisation ability of the trained models.

The networks in the datasets were generated by the Erdős-Rényi (ER) \cite{ER},  Watts-Strogatz (WS) model \cite{WS} and  Barabási-Albert (BA) models \cite{BA}. The parameters of these models were selected in such a way that the generated instances emphasize the characteristic features of each network family, while the other basic network statistics are kept about constant across the dataset. The characteristic features considered are: average path length ($\ell$), transitivity ($T$) and the shape of the degree distribution. For average path length and transitivity two cases were considered: low ($\ell<log(N)$ and $T<d$) and high (otherwise). Here $N$ denotes the number of nodes in the network, while $d$ stands for density (i.e., the fraction of all possible links present in the network). These thresholds mean that network instances with \textit{low} average path lengths have the so-called \textit{small-world property}, while  \textit{high} transitivity means that the value is higher than that expected in random networks. In the case of degree distribution, we also considered two cases: scale-free (SF)  and non-scale-free (NSF) distributions. 

There are 8 possible combinations of the \textit{high} and \textit{low} cases of these three properties. Since all generative models (ER, WS and BA) generate networks with \textit{low} $\ell$, we added regular lattices to cover the case of \textit{high} average path length. In these networks, nodes are placed on a virtual plane at equal distances both horizontally and vertically, such that they form a regular matrix. Edges are created between the nearest neighbors, but nodes on the edge of the matrix are also connected to the corresponding nodes at the opposing end (i.e., the matrix is 'wrapped around'). These periodic boundaries ensure that each node has exactly the same number of neighbors (and thus links). Two interpretations of neighborhood is used. The \textit{von Neumann} neighborhood includes the nodes one step to the north, east, south and west. The \textit{Moore} neighborhood adds the nodes in one step diagonally. These two interpretations yield distinctly different transitivity values:  low ($T=0$) in case the von Neumann neighborhood and high for Moore. This is useful for balancing our dataset. However, the two neighborhood interpretations introduce a variance, too. They yield 4 and 8 links per node, respectively. To counterbalance this, the parameters of the generative models are set such that they produce two kind of instances: with 4 and 8 links per node on average. 

Table \ref{tab:networks} summarizes the 8 network types contained in our synthetic datasets. They cover all four combinations for NSF degree distributions, at two levels of average degree. However, SF degree distributions are only covered by network instances generated by BA networks (at two levels of average degree) and all have low values for both $\ell$ and $T$. Generative models have been proposed  that produce networks with SF degree distributions and high clustering or high average path length. However, they are somewhat obscure (lesser known), therefore we decided to leave them out from the present version of the dataset.

We created two datasets with the above considerations. The \textbf{Small-Sized Graphs Dataset (Small Dataset)} contains 250 samples from each 8 network types and totals totalling at 2000 synthetic networks, with the size of the network ($N$) selected randomly from the $[250, 1024]$ interval. Table \ref{tab:SmallDSStats} summarizes the main network statistics of the Small Dataset. The \textbf{Medium-Sized Graphs Dataset (Medium Dataset)}
also contains 250 samples from each 8 network types, and thus totals at 2000 synthetic networks, but the networks' size ($N$) is now selected randomly from the $[1024, 2048]$ interval. 
During training the Small Dataset  was uniformly split into 1600 training graphs, 200 validation graphs, and 200 testing graphs, yielding a ratio of 80\%/10\%/10\%. The Medium Dataset serves as additional test data to test the generalising ability of the trained models.

For each network type, in both datasets,  the governing parameters were chosen to yield the desired average degree. For ER networks edge creation probability was set to $4/N$ and $8/N$ to yield an average degree of 4 and 8, respectively. In case of the WS model, the $k$ parameter (the number of connected nearest neighbors) was set to $4$ and $8$, respectively. Similarly, the number of links a new node connects to was set to $2$ and $4$ in the case of the BA model. In addition, in case of the WS model the probability of rewiring ($w$) was sampled uniformly from the range of $[0.1, 0.11]$ to ensure that the generated WS networks have both high transitivity and low average path length. Table \ref{tab:SmallDSStats} reports the main network statistics for the graphs in the Small Dataset.

\begin{table}[t!]
\centering
\caption{Summary of the main network statistics of the instances of the Small Dataset. The size of the network ($N=|V|$) is between 258 and 1022 with a mean of 644 for all network types.}  
\begin{tabular}{|l|c|c|c|c|c|c|c|}
\hline
& \multicolumn{5}{c|}{mean (min-max)}  \\
\hline
 Label  & \textbf{$|E|$} & \textbf{$\langle deg\rangle$} & \textbf{$d$} & \textbf{$T$} & \textbf{$\ell$} \\ \hline
 $ER_{low}$ & 1285 (528-2048) & 4.00 (3.58-4.40) & 0.01 (0.00-0-02) & 0.01 (0.00-0.03)& 4.51 (3.84-5.01) \\
 $ER_{high}$   & 2571 (1033-4152)& 7.99 (7.53-8.48) & 0.01 (0.01-0.03) & 0.01 (0.01-0.03)& 3.31 (2.88-3.63) \\
 $WS_{low}$ & 1287 (516-2044) & 4.00 (4.00-4.00)& 0.01 (0.00-0.02)) & 0.35 (0.31-0.38)& 7.85 (6.25-8.97)\\
 $WS_{high}$ & 2574 (1032-4088) & 8.00 (8.00-8.00) & 0.01 (0.01-0.03) & 0.46 (0.42-0.49)& 4.57 (3.72-5.11) \\
 $BA_{low}$  & 1283 (512-2040) & 3.99 (3.97-3.99) & 0.01 (0.00-0.02)& 0.02 (0.01-0.04) & 3.86 (3.30-4.17) \\
 $BA_{high}$ & 2558 (1016-4072)& 7.94 (7.88-7.97) & 0.01 (0.01-0.03)& 0.04 (0.02-0.07) & 3.01 (2.71-3.22) \\
 $GRID_{low}$  & 1287 (516-2044)& 4.00 (4.00-4.00) & 0.01 (0.00-0.02) & 0.00 (0.00-0.00) & 24.12 (8.27-64.56) \\
 $GRID_{high}$& 2574 (1032-4088)& 8.00 (8.00-8.00) & 0.01 (0.01-0.03)  & 0.43 (0.43-0.43) & 21.93 (5.63-63.57)  \\
 \hline
\end{tabular}
 \label{tab:SmallDSStats}
\end{table}

%-----------------------------------------------------------------------------------
\section{Results}

\begin{small}

\begin{table}[ht]
\centering
\caption{The generalisation ability of the various models. Rows correspond to GNN architectures, while columns for Augmentation Strategies. The three pairs of numbers in the cells are the minimum values of $H$ where 100\%, 95\% or 90\% accuracy was achieved, respectively -- in case of the Small (first number) and the Medium Dataset (second number). The last value is the models' generalisation ability at the 90\% accuracy level. }
\begin{tabular}{|c|c|c|c|c|c|}
\hline
%\begin{tabular}[c]{@{}c@{}}\\\\\\\end{tabular}
\textbf{} & \textbf{ID} & \textbf{Degree} & \textbf{Noise} & \textbf{Norm Degree} & \textbf{Ones}\\ \hline
\textbf{GIN} & \begin{tabular}[c]{@{}c@{}}8, 8\\2, 4\\2, 2\\100\%\end{tabular} & \begin{tabular}[c]{@{}c@{}}8, -\\8, 16\\8, 8\\100\%\end{tabular} & \begin{tabular}[c]{@{}c@{}}-, -\\8, 16\\4, 16\\50\%\end{tabular} & \begin{tabular}[c]{@{}c@{}}16, -\\8, -\\4, -\\ -\end{tabular} & \begin{tabular}[c]{@{}c@{}}8, -\\8, 8\\8, 8\\75\%\end{tabular}\\ \hline
\textbf {GATv2} & \begin{tabular}[c]{@{}c@{}}-, -\\4, 32\\4, 8\\75\%\end{tabular} & \begin{tabular}[c]{@{}c@{}}-, -\\8, 32\\8, 32\\33\%\end{tabular} & \begin{tabular}[c]{@{}c@{}}-, 16\\4, 8\\4, 4\\100\%\end{tabular} & \begin{tabular}[c]{@{}c@{}}-, -\\16, -\\16, -\\-\end{tabular} & \begin{tabular}[c]{@{}c@{}}-, -\\-, -\\-, -\\-\end{tabular} \\ \hline
\textbf{Global} & \begin{tabular}[c]{@{}c@{}}32, 32\\16, 32\\8, 8\\100\%\end{tabular} & \begin{tabular}[c]{@{}c@{}}-, -\\32, -\\16, 16\\50\%\end{tabular} & \begin{tabular}[c]{@{}c@{}}-, -\\32, -\\32, -\\-\end{tabular} & \begin{tabular}[c]{@{}c@{}}-, -\\16, -\\16, - \\-\end{tabular} & \begin{tabular}[c]{@{}c@{}}-, -\\-, -\\-, -\\-\end{tabular} \\ \hline
\textbf{Hierarchical} & \begin{tabular}[c]{@{}c@{}}-, -\\16, -\\16, -\\-\end{tabular} & \begin{tabular}[c]{@{}c@{}}32, -\\16, -\\8, -\\-\end{tabular} & \begin{tabular}[c]{@{}c@{}}-, -\\-, -\\-, -\\-\end{tabular} & \begin{tabular}[c]{@{}c@{}}32, -\\32, -\\32, -\\-\end{tabular} & \begin{tabular}[c]{@{}c@{}}-, -\\-, -\\-, -\\-\end{tabular}\\ \hline
\end{tabular}%
\label{tab:results-h-dependence}
\end{table}
\end{small}

\begin{figure}[b!]
\vspace{-0.8em}
    \centering
    \includegraphics[width=0.495\textwidth]{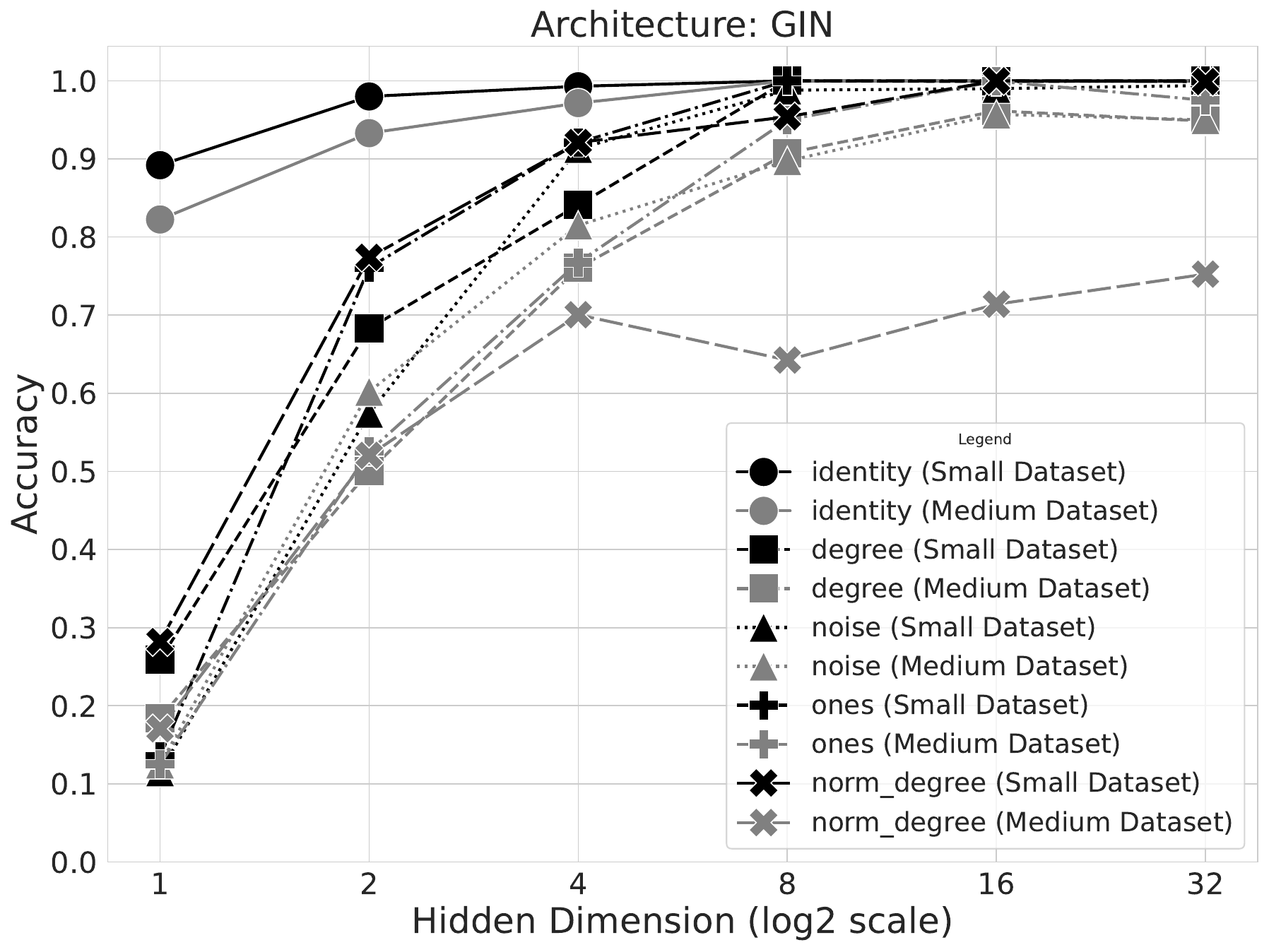}
    \includegraphics[width=0.495\textwidth] 
    {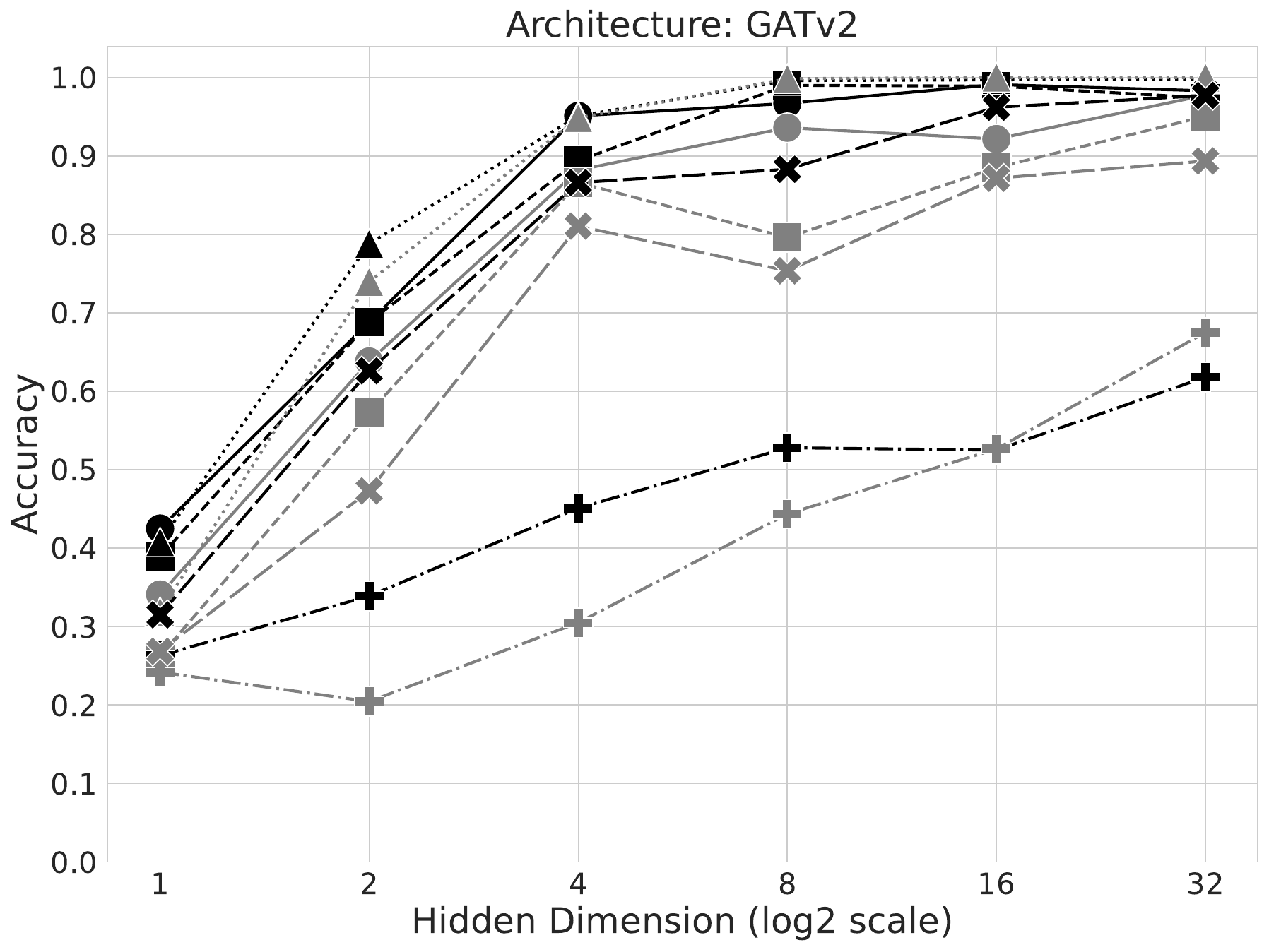}
    \includegraphics[width=0.495\textwidth]{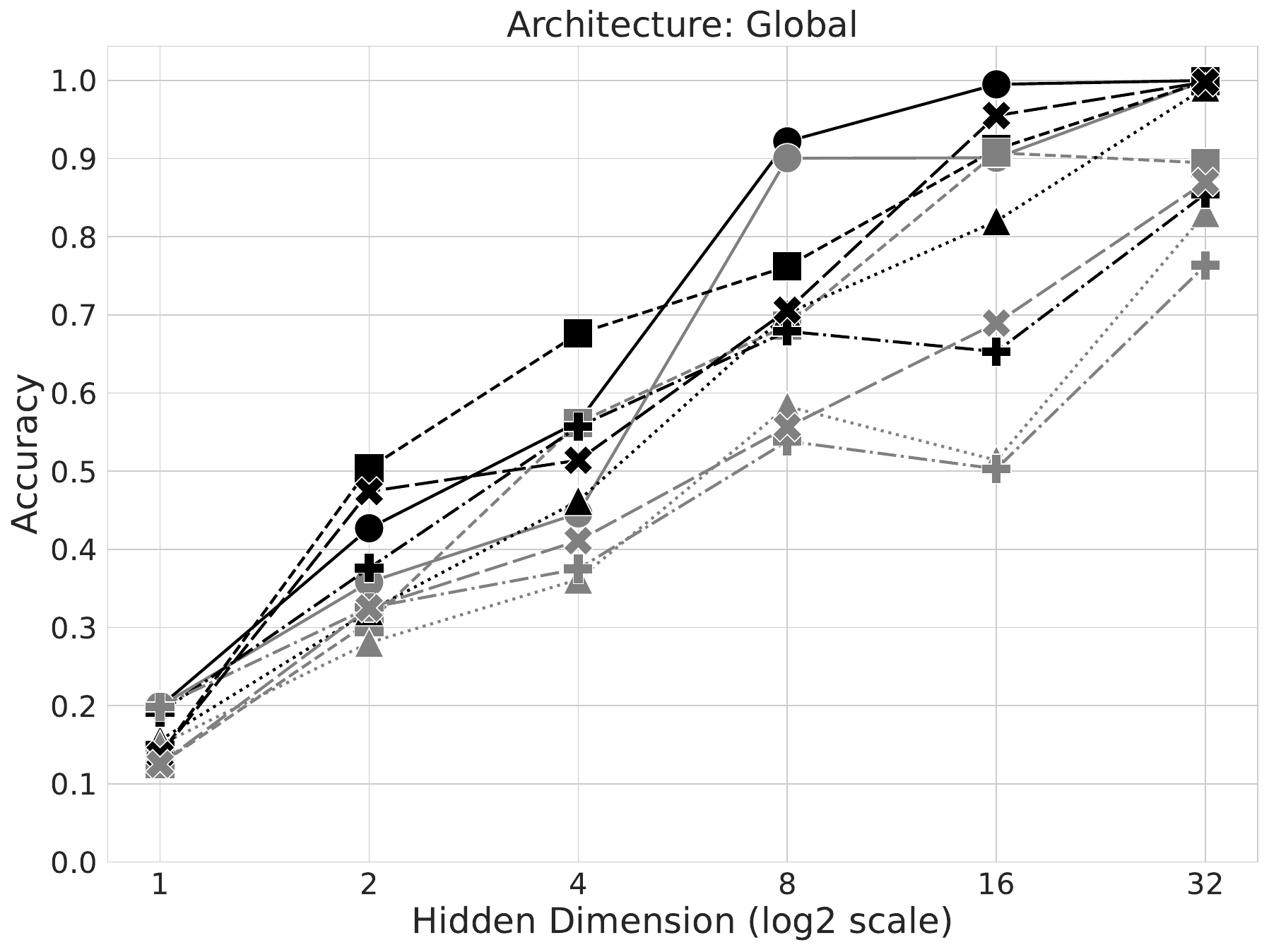}
    \includegraphics[width=0.495\textwidth]{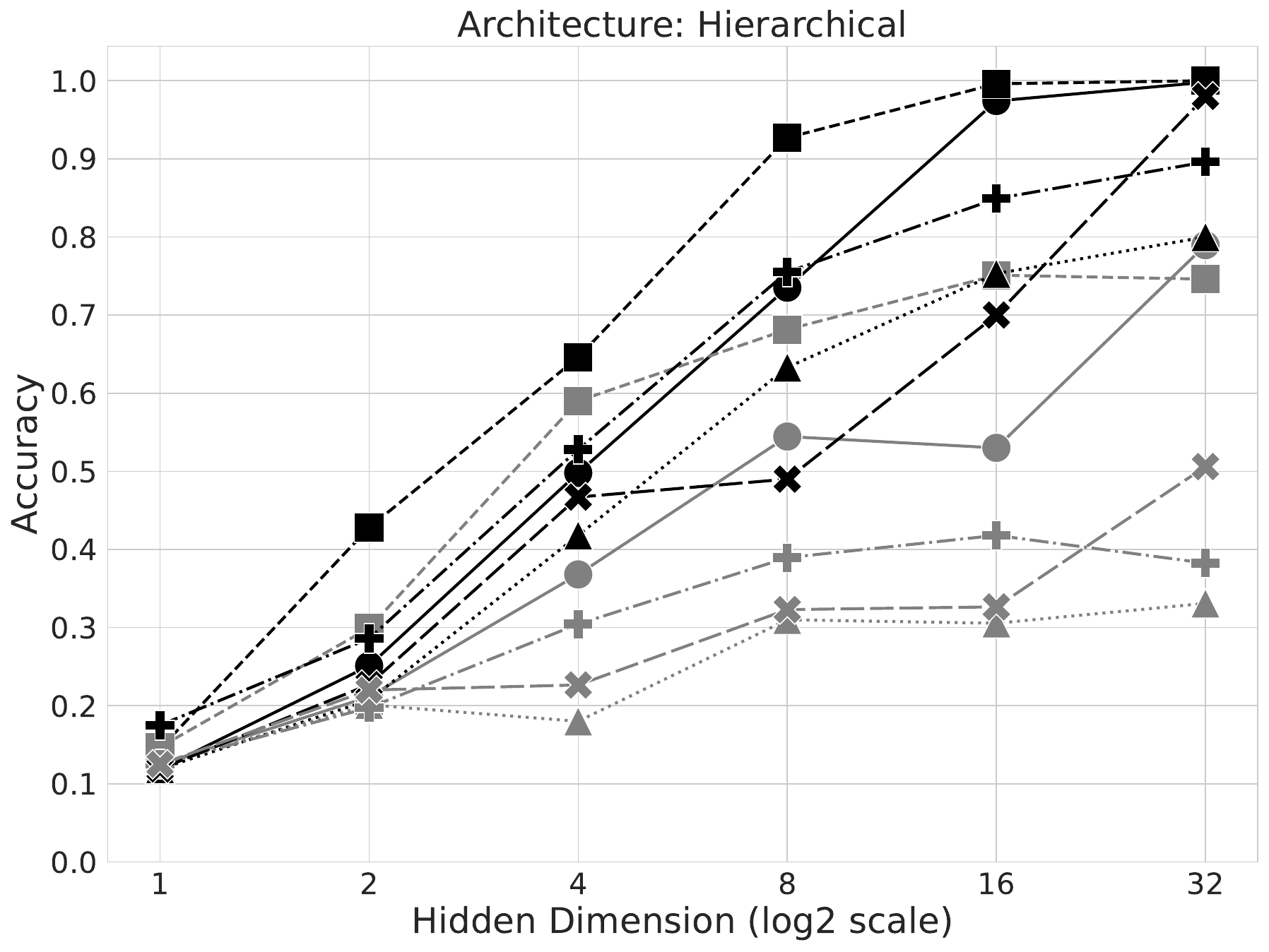}
    \caption{
%Comparison of Different Graph Neural Network Architectures Across Various Node Feature Types and Hidden Dimensions. Each subplot showcases the performance of different GNN architecture (GIN, GATv2, Hierarchical and Global) versus various hidden dimensions with different node features (identity, degree, ones, noise, norm degree). The GIN architecture, with its two linear layers, demonstrates exceptional generalization, even on medium-sized graphs. The GATv2 model with identity feature generalizes to medium-sized graphs, while noise features show robust generalization abilities. The HIERARCHICAL architecture with degree features achieves higher accuracy for small-sized graphs but fails to generalize for medium-sized graphs. The GLOBAL architecture with noise features shows escalating accuracy, but the identity feature consistently outperforms. Points on lines represent average accuracy from two trials with different weight initialization, tested on both small and medium-sized graphs. black markers denote accuracy on small-sized graphs, gray markers indicate accuracy on medium-sized graphs. (Hyperparameters: GIN Layers: 4, hierarchical Layers: 4, global Layers: 4, GATv2 Layers: 4; Batch Size: 100; Dropout: 0.5; Trials: 5; Optimizer: Adam; Epochs: 100; Weight Decay: 1e-3; Learning Rate: 0.01; GIN and global layers have two linear layers with batch normalization and ReLU activation, hierarchical layers have one linear layer, GATv2 layers have one linear layer.)
Performance of the various augmentation strategies and embedding architectures as a function of $H$ for both the testing subset of the Small Dataset (used for training) and for the Medium Dataset (to assess generalisation ability). The GIN and GATv2 architectures perform well for most feature types and also  generalise acceptably. The Global architecture only shows good accuracy at larger hidden dimensions and generalises only for some feature types. The Hierarchical architecture is poor both in terms of performance and generalisation.
The panels  report on the GIN (left) and GATv2 (right) architectures in the top row and on the Global (left) and Hierarchical (right) architectures at the bottom. The accuracies obtained for the Small Dataset are plotted in black, while the results for the Medium Dataset are marked with gray and dashed lines. Augmentation strategies are distinguished by different markers -- the same marker is used for the same feature type in case of both datasets. The accuracy values shown are averages over 5 trials with independent random weight initialisations, each obtained with $K=4$, a batch size of 100 and a dropout rate of $0.5$, using the ADAM optimizer for 100 epochs with a weight decay of $10^{-3}$ and a learning rate of 0.01.
}
    \label{fig:combined_plots}
\end{figure}

We studied all combinations of the 5 augmentation strategies and 4 GNN architectures. For each combination we performed a grid search of various hyperparameters ($H$, $K$ and depth of the generated ID features in case of the Identity augmentation strategy). 
%\hlc[cyan]{Environment seed fixed before training for consistent weight initialization. 5 trials per model controlled for performance variations. Parameters optimized to minimize loss function. Testing employed weights from the last epoch of the first trial, providing insights into model generalization on given datasets.}
Special attention was paid to hyperparameter $H$, or hidden dimension, which is the number of hidden nodes per layer in the basic graph convolution (message passing) steps of the GNN architectures. (For convenience, the same parameter also controls the number of hidden nodes in the classification head's MLP.) This parameter is important, as it has a tremendous impact on the computational requirements of the GNN models, both during training and prediction.

%\subsection{Performance of Various Augmentation Strategy and GCN Architecture Combinations}
%\hl{\textbf{WALID: please, regenerate the plots, such that the legend says Small Dataset and Medium Dataset, not small graphs and medium graphs. Also, everywhere in the text you use the term GATv2 instead of GAT. So, please, use the same in the plot(s) as well.}}
%\hlc[green]{\textbf{WALID: Please, redo the plots in such a way that the grids in the background are consistent. On the GAT plot the horizontal lines are with 0.1 distance between them, while on the others, these gaps are 0.2. I prefer the 0.1 step size.}}

Figure \ref{fig:combined_plots} summarizes the performance of the various augmentation strategies and embedding architectures as a function of $H$ for both the testing subset of the Small Dataset (used for training) and for the Medium Dataset (to assess generalisation ability). The panels  report on the GIN (left) and GATv2 (right) architectures in the top row and on the Global (left) and Hierarchical (right) architectures at the bottom. The accuracies obtained for the Small Dataset are plotted in black, while the results for the Medium Dataset are marked with gray and dashed lines. Augmentation strategies are distinguished by different markers -- the same marker is used for the same feature type in case of both datasets. The accuracy values shown are averages over 5 trials with independent random weight initialisations, each obtained with $K=4$, a batch size of 100 and a dropout rate of $0.5$, using the ADAM optimizer for 100 epochs with a weight decay of $10^{-3}$ and a learning rate of 0.01.

Table \ref{tab:results-h-dependence} provides a quantitative summary of the same results. The rows correspond to GNN architectures, while columns represent augmentation strategies. In each cell, there are 3 pairs of numbers, plus a percentage. The first pair reports the minimum value of $H$ at which 100\% accuracy was achieved for the Small and Medium Dataset, respectively. The second and third pair report the same information for the accuracy levels of 95\% and 90\%, respectively. If the given accuracy level was not reached at all, then '-' is used as a placeholder. Finally, the percentage summarises the given model's ability to generalise at the accuracy level of 90\%. It shows the ratio of above 0.9 accuracies for the Medium Dataset, over the number of 0.9 accuracies for the Small Dataset. If 0.9 accuracy was never reached for the Small Dataset, a '-' is used as a placeholder.

%Count how many times the model predicts above 90% for the small dataset (let’s call this A)
%and for the medium dataset (let’s call this B). Then show B/A as a percentage, if A isn’t 0. If it is, show ’-’.

%\hl{Explain -- averaged values, for the best hyperparameters
%}\hlc[red]{\textbf{Prof: I have added the describtion; WALID: Please, describe briefly how the hyperparameters are selected for the values shown on the plots.}}

%\hl{Overall picture: different features with different architectures. Shapes -- outliers.}

\textbf{The GIN architecture}, as the top left panel of Figure \ref{fig:combined_plots} shows, 
 works best with the ID (Identity) feature. All other features show very similar behavior, except for normalized degree in case of medium dataset. The models generalise well, in principle, showing a minor decrease in accuracy for the medium dataset. The one exception is the case of the normalized degree feature. Above 90\% accuracies are observed for $H\geq8$ for all feature types, but already at $H=2$ for ID and at $H=4$  for noise and normalized degree.
 %\hlc[red]{\textbf{Prof: I checked, and I did not switch between them, You can observe the behavior of the degree in the hierarchical architecture, as it was previously. I believe that running 5 trials altered the results. (Also, I'm training the norm degree again to update the plots based on the notes you provided about them.); WALID: Could you, please, check this? I remember, we had this behavior for the original degree feature. Then, we introduced the norm-degree to fix it. Are you sure you did not accidentally switched the labels?? }} Above 0.9 accuracies are observed from about $H=8$ (but already at $H=2$ for the ID feature).

In case of the \textbf{GATv2 architecture} (top right panel), the features generally behave similarly to each other (i.e., the plot is compact) and also to what was seen for the GIN architecture. Except for the case of the 'ones' feature, which performs poorly. Interestingly, the  ID feature, in contrast to the GIN architecture, is   outperformed or matched in many cases by simpler augmentation strategies (even by simple noise).  The models generalise well, in principle: there are relatively small differences between the results on the small and the medium dataset, but the generalisation errors are typically larger than in case of the GIN architecture. Above 0.9 accuracies are observed consistently from $H=4$ for the ID and noise features, from $H=8$ and $16$ in case of degree and normalized degree, respectively, and never for the 'ones' feature. 

The performance of the \textbf{Hieararchical architecture} is fairly poor in case of all feature types, compared to that of the GIN or GAT architectures. Generalisation is very poor for all feature types. Above 0.9 accuracies are only observed for the Small Dataset. Even there it only happens at high values of $H$ for most augmentation strategies (16 and beyond, except for the degree feature at $H=8$)  and never  in our studies for noise. On the other hand, the degree feature clearly outperforms the other feature types and this is true for the Middle Dataset as well. The normalized degree feature's performance scales with $H$ differently from the other augmentation strategies, showing an exponential-like trend as opposed to the saturation observed in other cases.

In case of the \textbf{Global architecture}: the behaviors of the 5 feature types are similar and  are generally fairly poor. Above 90\% accuracies are only observed at $H\geq16$ for most feature types (at $H=8$ for the ID feature and never for  'ones'). Generalisations (behavior on the Middle Dataset) is relatively poor for most augmentation strategies, but better than in the case of the Hierarchical architecture. Overall, the ID feature shows some advantage on the both datasets (with a slight drop at $H=16$ for the Middle Dataset).

%What hidden dimensions to work with in what case

\textbf{In summary}, the ID feature has the most consistent behavior across the 4 architectures. This corresponds to the level of structural information it contains about the network (i.e., cycle counts) that corresponds to a relatively high computational cost. The degree feature performs also comparably well across the architectures, which is consistent with its information content (with a much lower computational cost). %\hl{
Surprisingly, the normalized degree feature is not always able to generalise. %}  
Noise performs well for advanced architectures (GIN and GATv2), while it is generally poor in other cases. The 'ones' feature performs poorly, as expected (as it is not helpful in distinguishing nodes). Yet, the GIN architecture can make it work and it slightly beats (the also underperforming) noise feature in case of the Hierarchical architecture. The explanation of  the performance drop of the ID feature in case of the GATv2 model  at low values of $H$ (relative to the GIN model) requires further studies.

Generally speaking, the models were able to generalise to the Middle Dataset in case of the GIN architecture (except for the normalised degree  features). This corresponds to its increased computational ability (more hidden layers). The GATv2 architecture also shows generalisation ability for several feature types, but at lower levels of accuracy (and never for normalized degree or for the 'ones' feature). This is also congruent with the
more complex calculations it performs per message passing layer. In case of these architectures, acceptable performances and generalisations were possible also for low hidden dimensions: $H=2$ in case of the ID feature and $H=4, 8$ in case of others.
The Global architecture generalises perfectly with the ID feature and at 50\% level for the degree feature. However, it shows no generalisation ability for the other feature types. This shows that the architecture is able to use the structural information provided by the augmentation strategy, but it is unable to derive it from more elementary pieces of information.  
On the other hand, the Hierarchical architecture is not able to generalise at all.  

\section{Conclusions and Future Work}

This paper embarked on a study of various deep learning graph embedding approaches applied to the task of graph classification. To this end,  a dataset of synthetic networks was created, containing a balanced number of instances from classic generative network models with carefully chosen parameter combinations. This dataset was  then used to train a select set of graph classification models, using a variety of graph neural network approaches. The trained models were tested on both unseen samples from the original dataset and, to assess their ability to generalise, on synthetic networks produced using different parameter combinations for the generative network models.  
%The application of the best models to real-world social networks is also demonstrated.
In particular, four Graph Neural Network architectures (GNNs) were studied: GCN with Hierarchical and Global aggregation, GIN and GATv2.
Since synthetic networks do not contain (node or edge) features, five different augmentation strategies (artificial feature types) were applied to nodes. In order of increasing information content, these were: constant 1, random noise, degree, normalized degree and ID -- a vector of the number of cycles of various lengths. All combinations of the 4 GNNs and the 5 artificial feature types were studied as a function of the hidden dimension of artificial neural networks used in the GNNs. 

Our results point towards the balanced importance of the computational power of the GNN architecture and the the information level provided by the artificial features. GNN architectures with higher computational power, like GIN and GATv2, perform well for most augmentation strategies. On the other hand, artificial features with higher information content, like ID or degree, not only consistently outperform other augmentation strategies, but can also help GNN architectures with lower computational power to achieve good performance. The results of hyperparameter optimisation, generally consistent with out findings presented here, will be reported in subsequent papers, together with the results concerning the application of models trained on the synthetic dataset to the classification (i.e., analysis) of real-world networks.

\printbibliography

\end{document}